%% file: main.tex
\documentclass[conference]{IEEEtran}
\usepackage{balance}
\usepackage{float}
\input{settings}

\input{commands/commands.tex}
\IEEEoverridecommandlockouts
\begin{document}
\maketitle
\input{abstract}
\begin{IEEEkeywords}
Resilience, reconfigurable intelligent surface (RIS), intelligent reflecting surface (IRS), outage, failure, cell-free MIMO, resource allocation, quality of service.\vspace{-0.145cm}
\end{IEEEkeywords}
\thispagestyle{empty}
\pagestyle{empty}

\input{sections/introduction.tex}

\input{sections/system_model.tex}
\input{sections/problem_formulation.tex}

\input{sections/numerical_results.tex}

\input{sections/conclusion.tex}
\balance
\input{content/laststuff}

\footnotesize
\bibliographystyle{IEEEtran}
\bibliography{references}
\balance
\input{content/acronyms}
\balance
\end{document}

%% file: settings.tex
\usepackage{standard}
\usepackage{etex}

\usepackage{rotate}
\usepackage{algorithmic}
\usepackage{algorithm}
\usepackage{float}
\usepackage{tikz}
\usepackage{refcount}

\usepackage{pgf,tikz}
\usepackage{pgfplots}
\usepackage{pgfplotstable}
\usepackage{bm}
\usepackage{adjustbox}
\usetikzlibrary{calc}
\usepackage{relsize}
\usepackage{ifthen}

\usetikzlibrary{calc,fit,arrows,plotmarks,intersections,patterns,shapes,decorations,positioning, backgrounds, matrix}

\usepackage[numbers,sort&compress]{natbib}
\usepackage[ngerman, english]{babel}

\usetikzlibrary{arrows.meta}
\edef\wdArrowLength{2}
\tikzset{>={Latex[width=1.5mm,length=\wdArrowLength mm]}}

\usepackage{cancel}
\usepackage{mathtools}
\usepackage{shadethm}
\usepackage{empheq}
\usepackage{skull}
\usepgfplotslibrary{units}
\usepackage{calc}
\usepackage{nicefrac}
\usepackage{fancybox}
\usepackage{psfrag}
\usepackage{dsfont}

\title{RIS-enhanced Resilience in Cell-Free MIMO} 
\author{\IEEEauthorblockN{Kevin Weinberger\IEEEauthorrefmark{1}, Robert-Jeron Reifert\IEEEauthorrefmark{1}, Aydin Sezgin\IEEEauthorrefmark{1} and Ertugrul Basar\IEEEauthorrefmark{2}}
\IEEEauthorblockA{\IEEEauthorrefmark{1}Institute of Digital Communication Systems, Ruhr-University Bochum, Bochum, Germany\\
\IEEEauthorrefmark{2}Communications Research and Innovation Laboratory (CoreLab), Koç University, Sariyer, Turkey\\
Email: \{robert-.reifert,kevin.weinberger,aydin.sezgin\}@rub.de, ebasar@ku.edu.tr}
\thanks{This work was funded in part by the Federal Ministry of Education and Research (BMBF) of the Federal Republic of Germany (F\"orderkennzeichen 16KIS1235, MetaSEC).}
}
\date{\today}

\usepackage{graphicx}
\usepackage{placeins}

\usepackage{booktabs}
\usepackage{marvosym}
\usepackage{multirow}

\usepackage{latexsym, amsmath, amssymb, amsfonts, upgreek}

\usepackage[nolist]{acronym}

\usepackage{tikz}
\usetikzlibrary{calc,arrows,positioning,decorations,shadows,shapes,fadings,matrix}
\tikzset{>=latex'}
\tikzset{semithick}

\makeatletter
\providecommand{\IfElsePackageLoaded}[3]{\@ifpackageloaded{#1}{#2}{#3}}
\makeatother

\usepackage{subfigure}
\IfElsePackageLoaded{subfig}
	{\usepackage[subfigure]{tocloft}}{	
	\IfElsePackageLoaded{subfigure}
		{\usepackage[subfigure]{tocloft}}
		{\usepackage{tocloft}}
	}

\makeatletter

\def\tikz@delimiter#1#2#3#4#5#6#7#8{%
	\bgroup
		\pgfextra{\let\tikz@save@last@fig@name=\tikz@last@fig@name}%
		node[outer sep=0pt,inner sep=0pt,draw=none,fill=none,anchor=#1,at=(\tikz@last@fig@name.#2),#3]
		{%
			{\nullfont\pgf@process{\pgfpointdiff{\pgfpointanchor{\tikz@last@fig@name}{#4}}{\pgfpointanchor{\tikz@last@fig@name}{#5}}}}%
			\delimitershortfall\z@
			\resizebox*{!}{#8}{$\left#6\vcenter{\hrule height .5#8 depth .5#8 width0pt}\right#7$}%
		}
		\pgfextra{\global\let\tikz@last@fig@name=\tikz@save@last@fig@name}%
	\egroup%
}

%% file: commands/commands.tex
\newcommand{\skewedIRS}{
	\begin{tikzpicture}
		\node[rectangle,minimum width=1.5cm,minimum height=1.5cm,draw,yslant=-0.6,scale=0.5,fill=blue!20!green] at (0,0){};

		\node[rectangle,minimum width=0.05cm,minimum height=0.05cm,draw,yslant=-0.6,scale=0.25,fill=yellow!70!green] (base) at (-0.3,-0.12){};

         \foreach \y in {1,...,5}
         {
            \node[rectangle,minimum width=0.05cm,minimum height=0.05cm,draw,yslant=-0.6,scale=0.25,fill=yellow!70!green] at
            ($($(base)+(-31:0.14*0)$)+(90:0.12*\y)$){};
         }

      \foreach \x in {1,...,5}
      {
         \node[rectangle,minimum width=0.05cm,minimum height=0.05cm,draw,yslant=-0.6,scale=0.25,fill=yellow!70!green] at
         ($(base)+(-31:0.14*\x)$){};

         \foreach \y in {1,...,5}
         {
            \node[rectangle,minimum width=0.05cm,minimum height=0.05cm,draw,yslant=-0.6,scale=0.25,fill=yellow!70!green] at
            ($($(base)+(-31:0.14*\x)$)+(90:0.12*\y)$){};
         }
      }
	\end{tikzpicture}	
}
\tikzset{hexagon/.code={
	\draw (0,2) -- (-4,0) -- (0,-2) -- (4,0) -- (0,2);
}}

\tikzset{phone/.code={
   \node [rectangle,rounded corners=1.5pt,draw,minimum height=0.6cm, minimum width=0.35cm] at (0,0){};
   \node [rectangle,rounded corners=1.5pt,draw,minimum height=0.5cm, minimum width=0.3cm] at (0,0){};
}}

\makeatletter
\def\cantox@vector#1#2#3#4#5#6#7#8{%
  \dimen@.5\p@
  \setbox\z@\vbox{\boxmaxdepth.5\p@
   \hbox{\kern-1.2\p@\kern#1\dimen@$#7{#8}\m@th$}}%
  \ifx\canto@fil\hidewidth  \wd\z@\z@ \else \kern-#6\unitlength \fi
  \ooalign{%
    \canto@fil$\m@th \CancelColor
    \vcenter{\hbox{\dimen@#6\unitlength \kern\dimen@
      \multiply\dimen@#4\divide\dimen@#3 \vrule\@depth\dimen@\@width\z@
      \vector(#3,-#4){#5}%
    }}_{\raise-#2\dimen@\copy\z@\kern-\scriptspace}$%
    \canto@fil \cr
    \hfil \box\@tempboxa \kern\wd\z@ \hfil \cr}}
\def\bcancelto#1#2{\let\canto@vector\cantox@vector\cancelto{#1}{#2}}
\makeatother

\newcommand{\ifthen}[2]{\ifthenelse{#1}{#2}{}}

\newcommand{\IRS}{
	\begin{tikzpicture}
		\node[rectangle,minimum width=1.5cm,minimum height=1.5cm,draw,scale=0.5,fill=blue!20!green] at (0,0){};

		\node[rectangle,minimum width=0.05cm,minimum height=0.05cm,draw,scale=0.25,fill=yellow!70!green] (base) at (-0.3,-0.3){};

         \foreach \y in {1,...,5}
         {
            \node[rectangle,minimum width=0.05cm,minimum height=0.05cm,draw,scale=0.25,fill=yellow!70!green] at
            ($($(base)+(-31:0.14*0)$)+(90:0.12*\y)$){};
         }

      \foreach \x in {1,...,5}
      {
         \node[rectangle,minimum width=0.05cm,minimum height=0.05cm,draw,scale=0.25,fill=yellow!70!green] at
         ($(base)+(0:0.12*\x)$){};

         \foreach \y in {1,...,5}
         {
            \node[rectangle,minimum width=0.05cm,minimum height=0.05cm,draw,scale=0.25,fill=yellow!70!green] at
            ($($(base)+(0:0.12*\x)$)+(90:0.12*\y)$){};
         }
      }
	\end{tikzpicture}	
}

\newcommand{\listequationsname}{List of Formulas}
\newlistof{myequations}{equ}{\listequationsname}

\definecolor{myblue1}{rgb}{0,0,255}
\definecolor{myblue2}{rgb}{65,105,225}
\definecolor{myblue3}{rgb}{70,130,180}
\definecolor{myblue4}{rgb}{176,196,222}

\newcommand{\mytilde}{{\raise.17ex\hbox{$\scriptstyle\mathtt{\sim}$}}}
\newcommand{\naive}{}
\def\naive/{na\"{\i}ve}

\newcommand{\executeiffilenewer}[3]{%
\ifnum\pdfstrcmp{\pdffilemoddate{#1}}%
{\pdffilemoddate{#2}}>0%
{\immediate\write18{#3}}\fi%
}

\usepackage{tikz}
\usetikzlibrary{arrows}
\usetikzlibrary{patterns}
\tikzset{>=latex'}
\usetikzlibrary{shapes.geometric}
\usetikzlibrary{calc}
\usetikzlibrary{positioning, shapes,fadings,through}

\pgfkeys{/pgf/fpu}
\pgfmathparse{16383+1}
\edef\tmp{\pgfmathresult}
\pgfkeys{/pgf/fpu=false}
\newcommand{\iPhone}[3]{
	\coordinate (a) at (#1,#2);
	\draw [line width=0.25pt,rounded corners=(#3)*1mm,fill=white,scale=(#3)] (a)--($(a)+(0.67,0)$)--($(a)+(0.67,1.381)$)--($(a)+(0,1.381)$)--cycle;
	\draw [color=gray,line width=0.25pt,rounded corners=(#3)*0.8mm,fill=white,scale=(#3)] ($(a)+(0.015,0.015)$)--($(a)+(0.655,0.015)$)--($(a)+(0.655,1.366)$)--($(a)+(0.015,1.366)$)--cycle;
	\draw [line width=0.25pt,rounded corners=(#3)*0.04mm,scale=(#3)] ($(a)+(0.2875,1.266)$)--($(a)+(0.3825,1.266)$)--($(a)+(0.3825,1.281)$)--($(a)+(0.2875,1.281)$)--cycle;
	\draw[line width=0.25pt,scale=#3] ($(a)+(0.335,0.09)$) circle (0.055cm);
	\draw[line width=0.25pt,scale=#3] ($(a)+(0.335,0.09)$) circle (0.044cm);
	\draw[line width=0.25pt,scale=#3] ($(a)+(0.2275,1.2735)$) circle (0.015cm);
	\draw[line width=0.25pt,scale=#3] ($(a)+(0.335,1.32)$) circle (0.01cm);
	\draw [fill={rgb:black,1;white,4},line width=0.25pt,scale=(#3)] ($(a)+(0.042475,0.170195)$)--($(a)+(0.042475,0.170195)+(0.58505,0.0)$)--($(a)+(0.042475,0.170195)+(0.58505,1.04061)$)--($(a)+(0.042475,0.170195)+(0.0,1.04061)$)--cycle;
}
\newcommand{\basestation}[3]{
	\coordinate (a) at (#1,#2);
	\draw[line width=(#3)*1.5pt,scale=#3] ($(a)+(0, -1.08)$) -- ($(a)+(0, 1)$);
	\draw[line width=(#3)*1.5pt,scale=#3] ($(a)+(0,1)$) .. controls ($(a)+(-0.25,-0.5)$) .. ($(a)+(-0.5,-0.9)$);
	\draw[line width=(#3)*1.5pt,scale=#3] ($(a)+(0,1)$) .. controls ($(a)+(0.25,-0.5)$) .. ($(a)+(0.5,-0.9)$);
	\draw[line width=(#3)*1.5pt,scale=#3] ($(a)+(0, -0.9)$) -- ($(a)+(-0.35, -0.7)$);
	\draw[line width=(#3)*1.5pt,scale=#3] ($(a)+(0, -0.9)$) -- ($(a)+(0.35, -0.7)$);
	\draw[line width=(#3)*0.75pt,scale=#3] ($(a)+(-0.35, -0.65)$) -- ($(a)+(0, -0.5)$);
	\draw[line width=(#3)*0.75pt,scale=#3] ($(a)+(0.35, -0.65)$) -- ($(a)+(0, -0.5)$);
	\draw[line width=(#3)*1pt,scale=#3] ($(a)+(0, -0.6)$) -- ($(a)+(-0.3, -0.45)$);
	\draw[line width=(#3)*1pt,scale=#3] ($(a)+(0, -0.6)$) -- ($(a)+(0.3, -0.45)$);
	\draw[line width=(#3)*0.5pt,scale=#3] ($(a)+(-0.3, -0.45)$) -- ($(a)+(0, -0.32)$);
	\draw[line width=(#3)*0.5pt,scale=#3] ($(a)+(0.3, -0.45)$) -- ($(a)+(0, -0.32)$);
	\draw[line width=(#3)*0.75pt,scale=#3] ($(a)+(0, -0.3)$) -- ($(a)+(-0.22, -0.17)$);
	\draw[line width=(#3)*0.75pt,scale=#3] ($(a)+(0, -0.3)$) -- ($(a)+(0.22, -0.17)$);
	\draw[line width=(#3)*0.5pt,scale=#3] ($(a)+(-0.22, -0.17)$) -- ($(a)+(0, -0.07)$);
	\draw[line width=(#3)*0.5pt,scale=#3] ($(a)+(0.22, -0.17)$) -- ($(a)+(0, -0.07)$);;
	\draw[line width=(#3)*0.75pt,scale=#3] (a) -- ($(a)+(-0.18, 0.11)$);
	\draw[line width=(#3)*0.75pt,scale=#3] (a) -- ($(a)+(0.18, 0.11)$);
	\draw[line width=(#3)*0.5pt,scale=#3] ($(a)+(-0.18, 0.11)$) -- ($(a)+(0,0.2)$);
	\draw[line width=(#3)*0.5pt,scale=#3] ($(a)+(0.18, 0.11)$) -- ($(a)+(0,0.2)$);
	\draw[line width=(#3)*0.5pt,scale=#3] ($(a)+(0, 0.3)$) -- ($(a)+(-0.1, 0.37)$);
	\draw[line width=(#3)*0.5pt,scale=#3] ($(a)+(0, 0.3)$) -- ($(a)+(0.1, 0.37)$);
	\draw[line width=(#3)*0.25pt,scale=#3] ($(a)+(-0.1, 0.37)$) -- ($(a)+(0, 0.43)$);
	\draw[line width=(#3)*0.25pt,scale=#3] ($(a)+(0.1, 0.37)$) -- ($(a)+(0, 0.43)$);
	\draw[line width=(#3)*0.75pt,scale=#3] ($(a)+(0, 1.2)$) -- ($(a)+(0,1)$);
	\draw[fill=white,scale=#3] ($(a)+(0, 1.2)$) circle (0.05cm);
	\draw[line width=(#3)*1pt,decorate,decoration=expanding waves,decoration={segment length=(#3)*10pt},scale=#3] ($(a)+(0, 1.2)$) -- ($(a)+(0, 2.5)$);
}
\newcommand{\basestationtwoant}[3]{
	\coordinate (a) at (#1,#2);
	\draw[line width=(#3)*1.5pt,scale=#3] ($(a)+(0, -1.08)$) -- ($(a)+(0, 1)$);
	\draw[line width=(#3)*1.5pt,scale=#3] ($(a)+(0,1)$) .. controls ($(a)+(-0.25,-0.5)$) .. ($(a)+(-0.5,-0.9)$);
	\draw[line width=(#3)*1.5pt,scale=#3] ($(a)+(0,1)$) .. controls ($(a)+(0.25,-0.5)$) .. ($(a)+(0.5,-0.9)$);
	\draw[line width=(#3)*1.5pt,scale=#3] ($(a)+(0, -0.9)$) -- ($(a)+(-0.35, -0.7)$);
	\draw[line width=(#3)*1.5pt,scale=#3] ($(a)+(0, -0.9)$) -- ($(a)+(0.35, -0.7)$);
	\draw[line width=(#3)*0.75pt,scale=#3] ($(a)+(-0.35, -0.65)$) -- ($(a)+(0, -0.5)$);
	\draw[line width=(#3)*0.75pt,scale=#3] ($(a)+(0.35, -0.65)$) -- ($(a)+(0, -0.5)$);
	\draw[line width=(#3)*1pt,scale=#3] ($(a)+(0, -0.6)$) -- ($(a)+(-0.3, -0.45)$);
	\draw[line width=(#3)*1pt,scale=#3] ($(a)+(0, -0.6)$) -- ($(a)+(0.3, -0.45)$);
	\draw[line width=(#3)*0.5pt,scale=#3] ($(a)+(-0.3, -0.45)$) -- ($(a)+(0, -0.32)$);
	\draw[line width=(#3)*0.5pt,scale=#3] ($(a)+(0.3, -0.45)$) -- ($(a)+(0, -0.32)$);
	\draw[line width=(#3)*0.75pt,scale=#3] ($(a)+(0, -0.3)$) -- ($(a)+(-0.22, -0.17)$);
	\draw[line width=(#3)*0.75pt,scale=#3] ($(a)+(0, -0.3)$) -- ($(a)+(0.22, -0.17)$);
	\draw[line width=(#3)*0.5pt,scale=#3] ($(a)+(-0.22, -0.17)$) -- ($(a)+(0, -0.07)$);
	\draw[line width=(#3)*0.5pt,scale=#3] ($(a)+(0.22, -0.17)$) -- ($(a)+(0, -0.07)$);;
	\draw[line width=(#3)*0.75pt,scale=#3] (a) -- ($(a)+(-0.18, 0.11)$);
	\draw[line width=(#3)*0.75pt,scale=#3] (a) -- ($(a)+(0.18, 0.11)$);
	\draw[line width=(#3)*0.5pt,scale=#3] ($(a)+(-0.18, 0.11)$) -- ($(a)+(0,0.2)$);
	\draw[line width=(#3)*0.5pt,scale=#3] ($(a)+(0.18, 0.11)$) -- ($(a)+(0,0.2)$);
	\draw[line width=(#3)*0.5pt,scale=#3] ($(a)+(0, 0.3)$) -- ($(a)+(-0.1, 0.37)$);
	\draw[line width=(#3)*0.5pt,scale=#3] ($(a)+(0, 0.3)$) -- ($(a)+(0.1, 0.37)$);
	\draw[line width=(#3)*0.25pt,scale=#3] ($(a)+(-0.1, 0.37)$) -- ($(a)+(0, 0.43)$);
	\draw[line width=(#3)*0.25pt,scale=#3] ($(a)+(0.1, 0.37)$) -- ($(a)+(0, 0.43)$);
	\draw[line width=(#3)*1pt,decorate,decoration=expanding waves,decoration={segment length=(#3)*10pt},scale=#3] ($(a)+(0, 1.2)$) -- ($(a)+(0, 2.5)$);
	\draw[line width=(#3)*0.75pt,scale=#3] ($(a)+(0.2, 1)$) -- ($(a)+(0,1)$);
	\draw[line width=(#3)*0.75pt,scale=#3] ($(a)+(0.2, 1.2)$) -- ($(a)+(0.2,1)$);
	\draw[fill=white,scale=#3] ($(a)+(0.2, 1.2)$) circle (0.05cm);
	\draw[line width=(#3)*0.75pt,scale=#3] ($(a)+(-0.2, 1)$) -- ($(a)+(0,1)$);
	\draw[line width=(#3)*0.75pt,scale=#3] ($(a)+(-0.2, 1.2)$) -- ($(a)+(-0.2,1)$);
	\draw[fill=white,scale=#3] ($(a)+(-0.2, 1.2)$) circle (0.05cm);
}
\newcommand{\basestationtwoantno}[3]{
	\coordinate (a) at (#1,#2);
	\draw[line width=(#3)*1.5pt,scale=#3] ($(a)+(0, -1.08)$) -- ($(a)+(0, 1)$);
	\draw[line width=(#3)*1.5pt,scale=#3] ($(a)+(0,1)$) .. controls ($(a)+(-0.25,-0.5)$) .. ($(a)+(-0.5,-0.9)$);
	\draw[line width=(#3)*1.5pt,scale=#3] ($(a)+(0,1)$) .. controls ($(a)+(0.25,-0.5)$) .. ($(a)+(0.5,-0.9)$);
	\draw[line width=(#3)*1.5pt,scale=#3] ($(a)+(0, -0.9)$) -- ($(a)+(-0.35, -0.7)$);
	\draw[line width=(#3)*1.5pt,scale=#3] ($(a)+(0, -0.9)$) -- ($(a)+(0.35, -0.7)$);
	\draw[line width=(#3)*0.75pt,scale=#3] ($(a)+(-0.35, -0.65)$) -- ($(a)+(0, -0.5)$);
	\draw[line width=(#3)*0.75pt,scale=#3] ($(a)+(0.35, -0.65)$) -- ($(a)+(0, -0.5)$);
	\draw[line width=(#3)*1pt,scale=#3] ($(a)+(0, -0.6)$) -- ($(a)+(-0.3, -0.45)$);
	\draw[line width=(#3)*1pt,scale=#3] ($(a)+(0, -0.6)$) -- ($(a)+(0.3, -0.45)$);
	\draw[line width=(#3)*0.5pt,scale=#3] ($(a)+(-0.3, -0.45)$) -- ($(a)+(0, -0.32)$);
	\draw[line width=(#3)*0.5pt,scale=#3] ($(a)+(0.3, -0.45)$) -- ($(a)+(0, -0.32)$);
	\draw[line width=(#3)*0.75pt,scale=#3] ($(a)+(0, -0.3)$) -- ($(a)+(-0.22, -0.17)$);
	\draw[line width=(#3)*0.75pt,scale=#3] ($(a)+(0, -0.3)$) -- ($(a)+(0.22, -0.17)$);
	\draw[line width=(#3)*0.5pt,scale=#3] ($(a)+(-0.22, -0.17)$) -- ($(a)+(0, -0.07)$);
	\draw[line width=(#3)*0.5pt,scale=#3] ($(a)+(0.22, -0.17)$) -- ($(a)+(0, -0.07)$);;
	\draw[line width=(#3)*0.75pt,scale=#3] (a) -- ($(a)+(-0.18, 0.11)$);
	\draw[line width=(#3)*0.75pt,scale=#3] (a) -- ($(a)+(0.18, 0.11)$);
	\draw[line width=(#3)*0.5pt,scale=#3] ($(a)+(-0.18, 0.11)$) -- ($(a)+(0,0.2)$);
	\draw[line width=(#3)*0.5pt,scale=#3] ($(a)+(0.18, 0.11)$) -- ($(a)+(0,0.2)$);
	\draw[line width=(#3)*0.5pt,scale=#3] ($(a)+(0, 0.3)$) -- ($(a)+(-0.1, 0.37)$);
	\draw[line width=(#3)*0.5pt,scale=#3] ($(a)+(0, 0.3)$) -- ($(a)+(0.1, 0.37)$);
	\draw[line width=(#3)*0.25pt,scale=#3] ($(a)+(-0.1, 0.37)$) -- ($(a)+(0, 0.43)$);
	\draw[line width=(#3)*0.25pt,scale=#3] ($(a)+(0.1, 0.37)$) -- ($(a)+(0, 0.43)$);
	\draw[line width=(#3)*0.75pt,scale=#3] ($(a)+(0.2, 1)$) -- ($(a)+(0,1)$);
	\draw[line width=(#3)*0.75pt,scale=#3] ($(a)+(0.2, 1.2)$) -- ($(a)+(0.2,1)$);
	\draw[fill=white,scale=#3] ($(a)+(0.2, 1.2)$) circle (0.05cm);
	\draw[line width=(#3)*0.75pt,scale=#3] ($(a)+(-0.2, 1)$) -- ($(a)+(0,1)$);
	\draw[line width=(#3)*0.75pt,scale=#3] ($(a)+(-0.2, 1.2)$) -- ($(a)+(-0.2,1)$);
	\draw[fill=white,scale=#3] ($(a)+(-0.2, 1.2)$) circle (0.05cm);
}
\renewcommand{\skewedIRS}{
	\begin{tikzpicture}
		\node[rectangle,minimum width=1.5cm,minimum height=1.5cm,draw,yslant=-0.6,scale=0.5,fill=blue!20!green] at (0,0){};
		
		\node[rectangle,minimum width=0.05cm,minimum height=0.05cm,draw,yslant=-0.6,scale=0.25,fill=yellow!70!green] (base) at (-0.3,-0.12){};
		
		\foreach \y in {1,...,5}
		{
			\node[rectangle,minimum width=0.05cm,minimum height=0.05cm,draw,yslant=-0.6,scale=0.25,fill=yellow!70!green] at
			($($(base)+(-31:0.14*0)$)+(90:0.12*\y)$){};
		}
		
		\foreach \x in {1,...,5}
		{
			\node[rectangle,minimum width=0.05cm,minimum height=0.05cm,draw,yslant=-0.6,scale=0.25,fill=yellow!70!green] at
			($(base)+(-31:0.14*\x)$){};
			
			\foreach \y in {1,...,5}
			{
				\node[rectangle,minimum width=0.05cm,minimum height=0.05cm,draw,yslant=-0.6,scale=0.25,fill=yellow!70!green] at
				($($(base)+(-31:0.14*\x)$)+(90:0.12*\y)$){};
			}
		}
	\end{tikzpicture}	
}

\renewcommand{\IRS}{
	\begin{tikzpicture}
		\node[rectangle,minimum width=1.5cm,minimum height=1.5cm,draw,scale=0.5,fill=blue!20!green] at (0,0){};
		
		\node[rectangle,minimum width=0.05cm,minimum height=0.05cm,draw,scale=0.25,fill=yellow!70!green] (base) at (-0.3,-0.3){};
		
		\foreach \y in {1,...,5}
		{
			\node[rectangle,minimum width=0.05cm,minimum height=0.05cm,draw,scale=0.25,fill=yellow!70!green] at
			($($(base)+(-31:0.14*0)$)+(90:0.12*\y)$){};
		}
		
		\foreach \x in {1,...,5}
		{
			\node[rectangle,minimum width=0.05cm,minimum height=0.05cm,draw,scale=0.25,fill=yellow!70!green] at
			($(base)+(0:0.12*\x)$){};
			
			\foreach \y in {1,...,5}
			{
				\node[rectangle,minimum width=0.05cm,minimum height=0.05cm,draw,scale=0.25,fill=yellow!70!green] at
				($($(base)+(0:0.12*\x)$)+(90:0.12*\y)$){};
			}
		}
	\end{tikzpicture}	
}
\newcommand{\tiltedIRS}{
	\begin{tikzpicture}
		\node[rectangle,minimum width=1.5cm,minimum height=1.5cm,draw,scale=0.5,fill=blue!20!green] at (0,0){};
		
		\node[rectangle,minimum width=0.05cm,minimum height=0.05cm,draw,scale=0.25,fill=yellow!70!green] (base) at (-0.3,-0.3){};
		
		\foreach \y in {1,...,5}
		{
			\node[rectangle,minimum width=0.05cm,minimum height=0.05cm,draw,scale=0.25,fill=yellow!70!green] at
			($($(base)+(-31:0.14*0)$)+(90:0.12*\y)$){};
		}
		
		\foreach \x in {1,...,5}
		{
			\node[rectangle,minimum width=0.05cm,minimum height=0.05cm,draw,scale=0.25,fill=yellow!70!green] at
			($(base)+(0:0.12*\x)$){};
			
			\foreach \y in {1,...,5}
			{
				\node[rectangle,minimum width=0.05cm,minimum height=0.05cm,draw,scale=0.25,fill=yellow!70!green] at
				($($(base)+(0:0.12*\x)$)+(90:0.12*\y)$){};
			}
		}
	\end{tikzpicture}	
}

\newcommand{\sensor}[4]{
	\coordinate (a) at (#1,#2);
	\draw[#4,fill=#4,scale=#3,rounded corners=1.5pt] ($(a)+(-.4, -.1)$) rectangle ++(.8,-.3);
	\draw[#4,fill=#4,scale=#3] ($(a)+(-.4, 0)$) rectangle ++(.8,-.18);
	\draw[white,fill=white,scale=#3] ($(a)+(0, 0)$) circle (0.2cm);
	\draw[#4,fill=#4,scale=#3] ($(a)+(0, 0)$) circle (0.15cm);
	\draw[#4,line width=(#3)*1pt,decorate,decoration=expanding waves,decoration={segment length=(#3)*10pt},scale=#3] ($(a)+(0, 0)$) -- ($(a)+(0, 1.3)$);
}
\newcommand{\criticality}[4]{
	\coordinate (a) at (#1,#2);
	\draw[#4,scale=#3,rounded corners=1.5pt] ($(a)+(0,0)$) rectangle ++(.8,-.8);
	\node[#4,scale=#3,anchor=center,align=center] at ($(a)+(.14,-.15)$) {\LARGE\bfseries{!}};
	\draw[#4,scale=#3,rounded corners=1pt] ($(a)+(.05,-.7)$) -- ($(a)+(.4, -.05)$) -- ($(a)+(.75, -.7)$) -- cycle;
}
\newcommand{\actuator}[4]{
	\coordinate (a) at (#1,#2);
	\draw[#4,scale=#3] ($(a)+(-.2, .6)$) rectangle ++(.4,.41);
	\draw[#4,scale=#3] ($(a)+(0,0)$) -- ($(a)+(0, .6)$);
	\node[#4,anchor=center,align=center] at ($(a)+(0,.325)$) {\tiny A};
	\draw[#4,scale=#3] ($(a)+(0,0)$) -- ($(a)+(.9, .6)$) -- ($(a)+(.9, -.6)$) -- cycle;
	\draw[#4,scale=#3] ($(a)+(0,0)$) -- ($(a)+(-.9, .6)$) -- ($(a)+(-.9, -.6)$) -- cycle;
	\draw[#4,fill=#4!35,scale=#3] ($(a)+(0, 0)$) circle (0.3cm);	
}
\newcommand{\ap}[4]{
	\coordinate (a) at (#1,#2);
	\draw[white,fill=#4!40,scale=#3] ($(a)+(0, 0)$) circle (0.25cm);
	\draw[#4,fill=white,scale=#3] ($(a)+(0, 0)$) circle (0.15cm);
	\draw[#4,scale=#3] ($(a)+(0, 0)$) circle (0.4cm);
	\draw[#4,line width=(#3)*0.75pt,decorate,decoration=expanding waves,rotate=0,decoration={segment length=(#3)*5pt},scale=#3,yshift=-10pt] ($(a)+(.4, 0)$) -- ($(a)+(.8, 0)$);
	\draw[#4,line width=(#3)*0.75pt,decorate,decoration=expanding waves,rotate=0,decoration={segment length=(#3)*5pt},scale=#3,yshift=-10pt] ($(a)+(-.4, 0)$) -- ($(a)+(-.8, 0)$);
}

\usetikzlibrary{decorations}
\pgfdeclaredecoration{lightning bolt}{draw}{
	\state{draw}[width=\pgfdecoratedpathlength]{
		\pgfpathmoveto{\pgfpointorigin}%
		\pgfpathlineto{\pgfpoint{\pgfdecoratedpathlength*0.6}%
			{-\pgfdecoratedpathlength*.1}}%
		\pgfpathlineto{\pgfpoint{\pgfdecoratedpathlength*0.55}{0pt}}%
		\pgfpathlineto{\pgfpoint{\pgfdecoratedpathlength}{0pt}}%
		\pgfpathlineto{\pgfpoint{\pgfdecoratedpathlength*0.4}%
			{\pgfdecoratedpathlength*.1}}%
		\pgfpathlineto{\pgfpoint{\pgfdecoratedpathlength*0.45}{0pt}}%
		\pgfpathclose%
	}%
}

%% file: abstract.tex
\begin{abstract}
More and more applications that require high reliability and fault tolerance are realized with wireless network architectures and thus ultimately rely on the wireless channels, which can be subject to impairments and blockages. Hence, these architectures require a backup plan in the physical layer in order to guarantee functionality, especially when safety-relevant aspects are involved. To this end, this work proposes to utilize the reconfigurable intelligent surface (RIS) as a resilience mechanism to counteract outages. The advantages of RISs for such a purpose derive from their inherent addition of alternative channel links in combination with their reconfigurability. The major benefits are investigated in a cell-free multiple-input and
multiple-output (MIMO) setting, in which the direct channel paths are subject to blockages. An optimization problem is formulated that includes rate allocation with beamforming and phase shift configuration and is solved with a resilience-aware alternating optimization approach. Numerical results show that deploying even a randomly-configured RIS to a network reduces the performance degradation caused by blockages. This becomes even more pronounced in the optimized case, in which the RIS is able to potentially counteract the performance degradation entirely. Interestingly, adding more reflecting elements to the system brings an overall benefit for the resilience, even for time-sensitive systems, due to the contribution of the RIS reflections, even when unoptimized. \vspace{-0.015cm}
\end{abstract} 

%% file: sections/introduction.tex
\section{Introduction}
Fueled by the large-scale deployment of 5G systems, \ac{IoT} technologies connect huge numbers of low-power, low-complexity, and battery-limited devices, each of which requiring specific data throughputs \cite{ericsson}. More and more enterprises migrate the \ac{IoT} network architecture towards next-generation low-power wide-area access technologies, e.g., Cat-M and narrowband \ac{IoT}, which are more energy efficient, reliable, and enable higher capacities \cite{ericsson}. Critical \ac{IoT} (or mission-critical \ac{IoT}) has gained considerable attention in the recent years thanks to use-cases such as factory automation, remote monitoring/interaction, UAV control and vehicular communications \cite{crit1}, \cite{crit2}. Critical \ac{IoT} refers to \ac{IoT} applications that require high reliability and low latency, which lay the foundation to a plethora of operations that are dependent on continuous data streams.

\indent As a consequence, the role of the wireless communication network gains increased significance and bears a higher responsibility. However, this responsibility can become a problem, as the wireless channel is not only nondeterministic but also subject to shadowing and blockages, which might lead to outages because of failed packet deliveries. Depending on the situation, an outage can result in impacts of different kinds ranging from minor delays to safety-relevant aspects like harming the environment or on-site workers. Consequently, it becomes imperative that the wireless network has the ability to evaluate and react to outages, while maintaining an acceptable level of service \cite{resiliencemetric,RobertRes}.

One of the main challenges for such \ac{URLLC} systems stems from the inability of utilizing legacy retransmission-based methods in the next transmission block to account for failed packet deliveries. Instead, the system needs to continuously monitor the network and should be able to quickly apply mechanisms that counteract potential outage scenarios, thus offering resilience against failures \cite{resiliencemetric}. In the context of a wireless network, such \textit{resilience mechanisms} can be implemented by reallocating the network's resources accordingly \cite{RobertRes}. In this work, we propose the \ac{RIS} as a resilience mechanism, which is a metasurface comprised of multiple tunable reflecting elements that can introduce a phase shift to incoming signals in real-time \cite{howitworks,RIStut}. With this reconfigurability, an \ac{RIS} is able to support the wireless transmission in different ways, i.e., by extending coverage, suppressing interference, or changing the channel statistics \cite{SynBenefitsConf,SynBenefits,corrBj,BjonAtten}. As a resilience method, the advantages of \acp{RIS} are twofold: On the one hand, new \ac{RIS}-assisted paths are introduced to the system, which can be utilized in case of a blockage in the direct paths \cite{Basar1}. This extends the \textit{resilience scope} of the system, as the addition of the \ac{RIS} enables the recovery from situations (like blockages in all direct paths) that would previously result in failures. On the other hand, these new paths are customizable. Thus the adaptation to disruptions, that were in the resilience scope before deploying the RIS, can be improved upon by a smart configuration of the phase shifters.

To this end, this work investigates the fundamental concept of utilizing the \ac{RIS} as a resilience method in a cell-free \ac{MIMO} downlink system, in which a \ac{CP} serves single-antenna users through distributed \acp{AP}. In order to highlight the effects and influence of the \ac{RIS} more prominently, we assume that the \ac{CP} has perfect global instantaneous \ac{CSI}. Based on this information an optimization problem is formulated that jointly allocates the rates of the users, while designing the beamformers and phase shifters to minimize the network-wide adaption gap. To facilitate practical implementation, we propose a resilience-aware alternating optimization framework, which splits the given non-convex optimization problem into two convex sub-problems. Additionally, this framework is able to take the system-specific requirements of either a high-quality or quick recovery into account by determining a solution that satisfies this specific quality-time trade-off.

%% file: sections/system_model.tex
\section{System Model}\label{ch:Sysmod}
\begin{figure}
	\centering
   \centering
	\tikzset{every picture/.style={scale=0.9}, every node/.style={scale=0.9}}
	
	\definecolor{color1}{rgb}{0.0503832136000000,	0.0298028976000000,	0.527974883000000}%
	\definecolor{color2}{rgb}{0.493426212287426,	0.0115244183443711,	0.658032128750000}%
	\definecolor{color3}{rgb}{0.796382339000000,	0.277979917955297,	0.471322439808427}%
	\definecolor{color4}{rgb}{0.972230291500000,	0.581718982058214,	0.254084505027532}%
	\definecolor{color5}{rgb}{0.940015097000001,	0.975158357000000,	0.131325517000000}%
	\begin{tikzpicture}[scale=1.4]
		
		\draw [draw=none,fill=color3!40, postaction={path fading=north, fading angle=180, fill=color3!30},opacity=1] (1.2,1.5) ellipse (1.4cm and 0.8cm);
		\node[anchor=center,align=center] at (1.19,1.05){ \emph{blockage} area};
		\draw[draw=none, fill=white] (-.4,1.3) rectangle ++(.7,0.8);
		\draw[draw=color3, thick, pattern=north east lines, pattern color=color3] (-.4,1.3) rectangle ++(.7,0.8);
		\draw[draw=color3, fill=color3, thick] (-.39,1.3) rectangle ++(.1,-0.1);
		\draw[draw=color3, fill=color3, thick] (.19,1.3) rectangle ++(.1,-0.1);
		\draw[draw=none, fill=white] (2.0,1.5) rectangle ++(.7,0.8);
		\draw[draw=color3, thick, pattern=north east lines, pattern color=color3] (2.0,1.5) rectangle ++(.7,0.8);
		\draw[draw=color3, fill=color3, thick] (2.01,1.5) rectangle ++(.1,-0.1);
		\draw[draw=color3, fill=color3, thick] (2.59,1.5) rectangle ++(.1,-0.1);

		\node[]  at (1.6,3.4) {\tiltedIRS};
		\draw[fill=color2!30] plot[variable=\t,domain=0:360,smooth,samples=51,rotate=77,yshift=-.6cm,xshift=3.45cm]
		({180+18*sin(\t)}:{1.0*pow(sin(\t/2),15)+.3});
		\draw[fill=color2!30] plot[variable=\t,domain=0:360,smooth,samples=51,rotate=90,yshift=-1.575cm,xshift=3.2cm]
		({180+18*sin(\t)}:{1.2*pow(sin(\t/2),15)+.3});
		
		\node [] (ap1) at (-1.5,1.95) {\begin{tikzpicture}\ap{0}{0}{.5}{black};\end{tikzpicture}};
		\node [] (ap2) at (3.5,2.45) {\begin{tikzpicture}\ap{0}{0}{.5}{black};\end{tikzpicture}};
		\iPhone{.9}{1.55}{.2}
		\iPhone{1.5}{1.25}{.2}
		\iPhone{3.2}{1.1}{.2}
		\iPhone{3.95}{0.9}{.2}
		
		\draw[-latex, dotted, color3, line width=0.2mm] (ap1) to (-.4,1.8);
		\draw[-latex, dotted, color3, line width=0.2mm] (ap1) to (-.4,1.55);
		\draw[-latex, dashed, color2, line width=0.2mm] (ap1) to (1.2,3.2);
		\draw[-latex, dotted, color3, line width=0.2mm] (ap2) to (2.7,2.0);
		\draw[-latex, dotted, color3, line width=0.2mm] (ap2) to (2.7,1.75);
		\draw[-latex, dashed, color2, line width=0.2mm] (ap2) to (2,3.2);
		\draw[-latex, color1, line width=0.2mm] (ap2) to (3.35,1.4);
		\draw[-latex, color1, line width=0.2mm] (ap2) to (3.94,1.2);
		
		\node[] at (3.95,3.302) {\begin{tikzpicture}[scale=1.1,every node/.style={scale=1.1}]
						\draw[fill=white] (4.0365,-1.10) rectangle ++(2.35,-1.0);
						\draw[-latex, color1, line width=0.2mm] (4.1,-1.3) -- (4.6,-1.3);
						\node[anchor=west,align=left] at (4.5,-1.3) {\footnotesize Direct link};
						\draw[-latex, color2, dashed, line width=0.2mm] (4.1,-1.6) -- (4.6,-1.6);
						\node[anchor=west,align=left] at (4.5,-1.6) {\footnotesize RIS link};
						\draw[-latex, color3, dotted, line width=0.2mm] (4.1,-1.9) -- (4.6,-1.9);
						\node[anchor=west,align=left] at (4.5,-1.9) {\footnotesize Blocked link};\end{tikzpicture}};
				
				\node[] at (-.8,3.38) {\begin{tikzpicture}[scale=1.1,every node/.style={scale=1.1}]
								\draw[fill=white] (4.05,-1.1) rectangle ++(3.2,-0.8);
								\node at (4.29,-1.3) {\begin{tikzpicture}\iPhone{0}{1}{0.2};\end{tikzpicture}};
								\node[anchor=west,align=left] at (4.45,-1.3) {\footnotesize User};
								\node[scale=.3,/.style={scale=1.0}]  at (5.6,-1.3) {\tiltedIRS};
								\node[anchor=west,align=left] at (5.7,-1.3) {\footnotesize RIS};
								\node [] (ap1) at (4.3,-1.7) {\begin{tikzpicture}\ap{0}{0}{.3}{black};\end{tikzpicture}};
								\node[anchor=west,align=left] at (4.45,-1.7) {\footnotesize AP};
								\node[scale=.4] at (5.6,-1.7) {\begin{tikzpicture}\draw[draw=color3, thick, pattern=north east lines, pattern color=color3] (-.4,0) rectangle ++(.7,0.8);
										\draw[draw=color3, fill=color3, thick] (-.39,0) rectangle ++(.1,-0.1);
										\draw[draw=color3, fill=color3, thick] (.19,0) rectangle ++(.1,-0.1);\end{tikzpicture}};
								\node[anchor=west,align=left] at (5.7,-1.7) {\footnotesize Blocker};
						\end{tikzpicture}};
	\end{tikzpicture}
	\caption{\ac{RIS}-aided cell-free MIMO system circumventing outages, which would be caused a sudden blockage area due to mobile blockages.}
	\label{sys_mdl}
	\vspace{-0.325cm}
\end{figure}
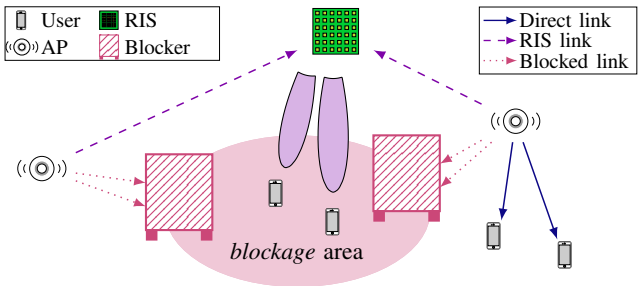%

This paper considers the \ac{RIS}-aided cell-free \ac{MIMO} downlink system depicted in Fig. 1. More precisely, a set of single-antenna users $\mathcal{K}=\{1,\dots,K\}$ is served by a set of $L$-antenna \acp{AP} $\mathcal{N}=\{1,\dots,N\}$. We consider the \ac{RIS} to be a uniform planar array, which is composed of $M$ passive reflecting elements. We assume that during the process of positioning the \ac{RIS}, it is assured that it is able to provide an alternative path for every user in case of blockages in their direct links. The \acp{AP}, as well as the \ac{RIS}, are connected to the \ac{CP} via perfect orthogonal fronthaul links, facilitating central processing at the \ac{CP}. In addition, each user has a \ac{QoS} target represented by a desired data rate $r_k^\mathsf{des}$.
\subsection{Channel Model}\label{ssec:chan}
	The channel model considered in this paper assumes quasi-static block fading channels, where the channel coefficients remain constant within the coherence time, but may change independently among coherence blocks. We denote the direct channel link between \ac{AP} $n$ and user $k$ as $\vect{h}_{n,k} \in \mathbb{C}^{L\times1}$. The reflected-channel link provided by the \ac{RIS} between \ac{AP} $n$ and user $k$ is denoted by $\mat{G}_{n,k} = \mat{H}_{n}\mathsf{diag}(\vect{v})\vect{g}_k \in \mathbb{C}^{L \times 1} $, where $\mat{H}_{n} \in \mathbb{C}^{L\times M}$ denotes the link between \ac{AP} $n$ and the \ac{RIS}, $\vect{g}_k \in \mathbb{C}^{M\times 1}$ denotes the link between the \ac{RIS} and user $k$ and $\mathsf{diag}(\vect{v})\in \mathbb{C}^{M \times M}$ denotes the reflection coefficient matrix with $\vect{v} = [v_1,v_2,\dots v_M] \in \mathbb{C}^{M \times 1}$. Here, $v_m = e^{j\theta_m}$ is the reconfigurable reflection coefficient at the $m$-th reflecting element, which is composed of a phase shift $\theta_m \in [0,2\pi]$. Further, we denote the aggregate direct channel vector of user $k$ as $\vect{h}_k= [\vect{h}_{1,k}^T,\vect{h}_{2,k}^T,\dots,\vect{h}_{N,k}^T]^T \in \mathbb{C}^{NL\times 1}$, the aggregate \ac{AP} to \ac{RIS} channel matrix as $\vect{H}=[\vect{H}_{1}^T,\vect{H}_{2}^T,\dots,\vect{H}_{N}^T]^T \in \mathbb{C}^{NL \times M}$ and the aggregate transmit signal vector as $\vect{x}=[\vect{x}_1^T,\vect{x}_2^T,\dots, \vect{x}_N^T]^T \in \mathbb{C}^{NL \times 1} $.

Utilizing the aggregate vectors, the received signal at user $k$ can be expressed as the sum of the direct and reflected channel vectors, namely\vspace{-0.16cm}
\begin{align}\label{recSgn}
  y_k =& (\vect{h}_k + \mat{G}_k\vect{v})^H \vect{x} + n_k,
\end{align}
where $\mat{G}_k=\mat{H}\text{diag}(\vect{g}_k)$ and $n_k\sim\mathcal{C}\mathcal{N}(0,\sigma_k)$ is the \ac{AWGN} sample.

Further, the symbols intended to be decoded by user $k$ are denoted by $s_k$. We assume that these messages form an \ac{i.i.d.} Gaussian codebook. These symbols for user $k$ are transmitted by the $n$-th \ac{AP} using the beamforming vector $\vect{w}_{n,k}\in\mathbb{C}^{L\times 1}$, which are both provided by the \ac{CP} over an ideal fronthaul.

\newcommand{\RisChan}[1]{\vect{h}_{#1} + \mat{G}_{#1}\vect{v}}

Hence, the overall transmit signal vector at the $n$-th \ac{AP} is given as $\vect{x}_{n}=\sum_{k\in\mathcal{K}} \vect{w}_{n,k} s_k$, which is subject to the power constraint $\mathbb{E}\{\vect{x}_n^H \vect{x}_n\} \leq P^{\mathsf{Max}}_n$, or equivalently \vspace{-0.05cm}
\begin{align} \label{powConst}
 \sum_{k\in\mathcal{K}} \vect{w}_{n,k} \leq P^{\mathsf{Max}}_n , \forall n \in \mathcal{N}.
\end{align}
The received signal (\ref{recSgn}) at user $k$ is then given by \vspace{-0.05cm}
\begin{equation}
    y_k = (\RisChan{k})^H \vect{w}_k s_k + \sum_{i\in\mathcal{K}\setminus\{k\}} (\RisChan{k}) \vect{w}_i s_i + n_k,\label{eq:yk}
\end{equation}
where the first term is the desired signal at the user and the second term is the received interference from all other users. Thus, we formulate the \ac{SINR} of user $k$ decoding its message as
\begin{align}\label{SINR}
	\Gamma_k = \frac{|(\RisChan{k})^H \vect{w}_k|^2}{\sum_{i\in\mathcal{K}\setminus\{k\}}|(\RisChan{k})^H \vect{w}_i^H|^2 + \sigma^2}.
\end{align}
Using these definitions, the \ac{QoS} demands for each user are satisfied, if the following conditions are met
\begin{align}
	r_k^\mathsf{des} \leq r_k \leq B \log_2(1+\Gamma_k) ,\quad \forall k \in \mathcal{K},
\end{align}
where $B$ denotes the transmission bandwidth and $r_k$ denotes the rate of user $k$.

%% file: sections/problem_formulation.tex
\subsection{Resilience Metric}
This paper considers a specific target throughput, e.g., determined by the \ac{QoS} requirements by the network with $r_k^\mathsf{des}$ being user $k$'s desired \ac{QoS} requirement. Note that the \ac{QoS} demands are assumed to remain constant within the observation interval. In contrast, the network's sum throughput is dependent on the allocated resources at time $t$, and thus captured in $\sum_{k=1}^K r_k(t)$, where $r_k(t)$ is the allocated data rate for user $k$ at time $t$. Regarding the time axis, there are two major cornerstones on the resilience behavior, namely $t_0$, the initial time at which the degradation manifests and $t_n$, the time of recovery. With those aspects at hand, and considering the resilience metric proposed in	\cite{resiliencemetric,RobertRes}, we define the networks \emph{absorption}, \emph{adaption}, and \emph{time-to-recovery} metrics as
	\begin{align}
	    r_\text{abs} &=  \frac{1}{K} \sum_{k\in\mathcal{K}} \frac{r_k(t_0)}{r_k^\mathsf{des}},\label{eq:rabs}\,\quad r_\text{ada} = \frac{1}{K} \sum_{k\in\mathcal{K}} \frac{r_k(t_n)}{r_k^\mathsf{des}},\quad
\end{align}\vspace{-0.3cm}
	\begin{align}
	    &\quad \quad r_\text{rec} =  \begin{cases} 1 \qquad\quad t_n-t_0\leq T_0\\  \frac{T_0}{t_n-t_0} \quad\, \text{otherwise} \end{cases}\hspace{-0.3cm}, \quad\quad\label{eq:rrec}
	\end{align}
	respectively, where $T_0$ is the network operator's desired recovery time, i.e., the time for which a functionality degradation is tolerable. A linear combination of the equations (\ref{eq:rabs})-(\ref{eq:rrec}) yields the considered resilience metric \vspace{-0.05cm}
	\begin{align}
	    r = \lambda_1 r_\text{abs} + \lambda_2 r_\text{ada} + \lambda_3 r_\text{rec},\label{eq:r}\\[-15pt]\nonumber
	\end{align}
	with fixed weights $\lambda_i$, $i\in\{1,2,3\}$, denoting the network operator's needs, e.g., emphasizing the robustness, the quality of adaption, or the recovery time. We also let $\sum_{i=1}^{3}\lambda_i = 1$, thus, the best-case value for the resilience is $r=1$.

Note that the proposed resilience metric (\ref{eq:r}) can be temporally divided in \textit{anticipatory} actions, i.e., $r_\text{abs}$, that take place before the outage occurs and \textit{reactionary} actions , i.e., $r_\text{ada}$ and $r_\text{rec}$, which occur after the outage. Regarding the anticipatory actions, various work are concerned with designing wireless communication networks to be robust against adverse conditions, e.g., \cite{scalableRS}. However, the literature on designing resilient wireless communication systems from the physical layer resource management perspective with a focus on adaption and time-to-recovery remains limited in breadth and depth. Thus, this paper mainly focuses the reactionary actions as they are sufficient to demonstrate the trade-off between the quality of a solution and the time necessary to obtain it. Consequently, we assume $\lambda_1=0$ throughout this work.

	\section{Problem Formulation}
	In order to study the effect of the \ac{RIS} on the resilience performance, we consider the problem of minimizing the constrained network-wide adaption gap, namely
	\begin{align}\label{Prob1}
		\underset{\vect{w},\vect{v},\vect{r}}{\min} \quad & \Uppsi =  \sum_{k\in \mathcal{K}} \Big|\frac{r_k}{r_k^\mathsf{des}}-1\Big| \tag{P1} \\[-5pt]
        \text{s.t.}\quad \,\,& (\ref{powConst}), \nonumber \\
        			&r_k \leq B \log_2(1+\Gamma_k) , \, \hspace{0.01cm} \forall k \in \mathcal{K}, \label{rateConst}\\
        			&|v_m| = 1 ,\quad \quad \quad \quad \,\,\,\,\,\, \, \forall m \in \{1,\dots,M\}, \label{unit_mod} \\[-15pt]\nonumber
	\end{align}
where $\vect{r} = [r_1, r_2, \dots, r_K]^T$ is the stacked rate vector and (\ref{unit_mod}) are the unit modulus constraints representing the phase shift constraints $0\leq \theta_m \leq 2\pi, \forall m \in \{1,\dots,M\}$. It can be observed that the feasible set of the formulated problem is non-convex  due to the non-convex nature of its constraints (\ref{rateConst}), (\ref{unit_mod}) and strong coupling of variables in (\ref{rateConst}).
	Therefore, we solve the problem by decoupling the variables with the alternating optimization approach proposed in \cite{SynBenefits}, where both emerging sub-problems are efficiently solved using the same \ac{SCA} framework. In addition, these sub-problems can also be considered as standalone resilience mechanisms (see Algorithm 1), facilitating a more in-depth study of the impact of the \ac{RIS} on the system's resilience.
	\subsection{Beamforming Design}
	As a result of the alternating optimization approach, the phase shifters are assumed to be fixed for the duration of the beamforming design. Thus, problem (\ref{Prob1}) can be rewritten as
\begin{align}\label{Prob2}
		\underset{\vect{w},\vect{r},\vect{q}}{\min} \quad & \Uppsi \quad  \quad \quad \tag{P2} \\[-5pt]
		\text{s.t.}\quad \,\, & (\ref{powConst}),\nonumber \\[-1pt]
   &r_k \leq B \log_2(1+q_k) , \, \,\forall k \in \mathcal{K}, \label{P2rate}
\end{align}
\begin{align}
		&q_k \leq \Gamma_k , \, \hspace{1.835cm} \forall k \in \mathcal{K}, \label{P2SINR}\\
		&\vect{q} \geq   0   \label{P2t},
	\end{align}
	where the introduction of the slack variables $\vect{q}=[q_1 ,\dots ,q_K]$ convexifies the rate expressions and $\vect{q} \geq 0$ signifies an element-wise inequality.  However, the constraints in (\ref{P2SINR}) are still non-convex but can be convexified using the \ac{SCA} approach. To this end, we rewrite (\ref{P2SINR}) as
	\begin{align}\label{SINR_rewritten}
		\sum_{i\in\mathcal{K}\setminus\{k\}}\hspace{-.2cm}|(\RisChan{k})^H \vect{w}_i|^2 + \sigma^2 - \frac{|(\RisChan{k})^H \vect{w}_k|^2}{q_k}\leq 0,
	\end{align}
	and apply the first-order Taylor approximation around the point $(\tilde{\vect{w}},\tilde{\vect{q}})$ on the fractional term. Consequently, the following convex approximation of (\ref{SINR_rewritten}) can be derived, see also \cite{SynBenefits},
	\begin{align}\label{SINR_convex}
	\sum_{i\in\mathcal{K}\setminus\{k\}}\hspace{-.2cm}|(\RisChan{k})^H \vect{w}_i|^2 + \sigma^2 +
			\frac{|(\RisChan{k})^H \tilde{\vect{w}}_k|^2}{(\tilde{q}_k)^2}q_k \nonumber \\ - \frac{2 \text{Re} \{\tilde{\vect{w}}_k^H(\RisChan{k})(\RisChan{k})^H\vect{w}_k \}}{\tilde{q_k}} \leq 0, \forall k \in \mathcal{K}.
	\end{align}
Thus, the approximation of problem (\ref{Prob2}) can be written as
\begin{align}\label{Prob2.1}
	\underset{\vect{w},\vect{r},\vect{q}}{\min} \quad & \Uppsi  \tag{P2.1} \\
	\text{s.t.}\quad \,\, & (\ref{powConst}),(\ref{P2rate}),(\ref{P2t}), (\ref{SINR_convex})\nonumber.
\end{align}
Problem (\ref{Prob2.1}) is convex and can be solved iteratively using the \ac{SCA} method. More precisely, we define $\mat{\Lambda}_z^w = [\vect{w}_z^T ,\vect{\kappa}_z^T ]^T$ as a vector stacking the optimization variables of the beamforming design problem at iteration $z$, where $\vect{\kappa}_z = [\vect{r}_z^T,\vect{q}_z^T]^T$. Similarly $\hat{\mat{\Lambda}}_z^w  = [\hat{\vect{w}}_z^T ,\hat{\vect{\kappa}}_z^T ]^T$ and $\tilde{\mat{\Lambda}}_z^w  = [\tilde{\vect{w}}_z^T ,\tilde{\vect{\kappa}}_z^T ]^T$ denote the optimal solutions and the point, around which the approximations are computed, respectively. Thus, with a given point $\tilde{\mat{\Lambda}}_{z}^w $, an optimal solution $\hat{\mat{\Lambda}}_z^w $ can be obtained by solving problem (\ref{Prob2.1}).

\newcommand{\RisChanv}[2]{\tilde{h}_{#1,#2} + \tilde{\mat{G}}_{#1,#2}\vect{v}}

\subsection{Phase Shift Design}
During the design process of the phase shifters at the \ac{RIS}, the beamformers are assumed to be fixed due to the application of the alternating optimization approach. With the intent of utilizing a similar problem structure as in (\ref{Prob2}), we denote $|(\RisChan{i})^H \vect{w}_k|^2$ $=$ $\tilde{h}_{i,k} + \tilde{\mat{G}}_{i,k} \vect{v}$, where $\tilde{h}_{i,k} = \vect{w}_k^H \vect{h}_i$ and $\tilde{\mat{G}}_{i,k} = \vect{w}_k^H \mat{G}_i$. With the above definitions, the \ac{SINR} constraints can be written similar to (\ref{SINR_rewritten}) as
\begin{align}\label{SINR_rewritten_v}
	\sum_{i\in\mathcal{K}\setminus\{k\}}\hspace{-.2cm}|\RisChanv{k}{i}|^2 + \sigma^2 - \frac{|(\RisChanv{k}{k})|^2}{q_k}\leq 0.
\end{align}
At this point, the overall optimization problem for the phase shift design can be formulated as
\begin{align}\label{Prob3}
	\underset{\vect{v},\vect{r},\vect{t}}{\min} \quad & \Uppsi + C\sum_{m=1}^{M}(|v_m|^2 -1) \tag{P3} \\
	\text{s.t.} \quad \, &(\ref{P2rate}),(\ref{P2t}),({\ref{SINR_rewritten_v}}), \nonumber
\end{align}
where the penalty method \cite{PenaltyMethod} for the phase shift constraints (\ref{unit_mod}) is adopted and $C$ is a large positive constant. Similar to (\ref{SINR_rewritten}), (\ref{SINR_rewritten_v}) can be approximated by calculating the first-order Taylor approximation of (\ref{SINR_rewritten_v}) on $\vect{v}$ around the point $(\tilde{\vect{v}},\tilde{\vect{q}})$, denoted by $(\widetilde{\ref{SINR_rewritten_v}})$. Further, the objective function can be approximated around the point $\tilde{\vect{v}}$ by first-order Taylor approximation of the penalty term $C\sum_{m=1}^{M}(|v_m|^2 -1)$, which is given by $C\sum_{m=1}^{M}\text{Re}\{2\tilde{v}_m^*v_m-|\tilde{v}_m|^2\}$. Based on the above approximation methods, problem (\ref{Prob3}) is approximated by the following convex problem:
\begin{align}\label{Prob3.1}
	\underset{\vect{v},\vect{r},\vect{t}}{\min} \quad & \Uppsi + C\sum_{m=1}^{M}\text{Re}\{2\tilde{v}_m^*v_m-|\tilde{v}_m|^2\} \tag{P3.1} \\
	\text{s.t.}  \quad \, &(\ref{P2rate}),(\ref{P2t}),(\widetilde{\ref{SINR_rewritten_v}}). \nonumber
\end{align}
Due to the similarity of the problem formulation and the utilization of the same \ac{SCA} framework, problem (\ref{Prob3.1}) can be solved by defining $\mat{\Lambda}_z^v = [\vect{v}_z^T, \vect{\kappa}_z^T]^T$ and following the same iterative procedure as for solving problem (\ref{Prob2.1}).

\subsection{Resilience-aware Alternating Optimization}
In this section, we outline an alternating optimization procedure, that is suitable to be utilized for resilience applications. Usually, when considering an \ac{SCA} approach in the literature, the goal is to retrieve the highest-quality solution for a given problem \cite{SynBenefits,PenaltyMethod,SynBenefitsConf}. Hence, these works do not take the duration of obtaining these solutions into consideration and employ time-intensive outer and inner loops, until some convergence criteria are met.

In the context of resilience, however, we are not necessarily interested in the highest-quality results. Instead we aim to obtain a solution, which satisfies a specific quality-time trade-off specified by the weights $\lambda_i$ in (\ref{eq:r}). Consequently, the proposed resilience-aware alternating optimization works towards minimizing the network wide gap $\Uppsi$ and stops as soon as it lies below a certain threshold $\tau$. Further, it dispenses from the convergence criteria of the inner loops when solving the sub-problems. Instead, a fixed amount of iterations $T_o, \forall o \in \{w,v\}$ of solving the problems (\ref{Prob2.1}) and (\ref{Prob3.1}) are introduced. This is possible due to the utilization of the same framework for both sub-problems because it makes $\tilde{\vect{\kappa}}$ a feasible point for both problems without additional adaption. The advantage of using fixed iterations lies in the fact that the intermediate solutions $\hat{\vect{w}}_z$ and $\hat{\vect{v}}_z$ of these sub-problems are also suitable to be evaluated in order to improve the resilience metric. This enables the algorithm to react to different values of $\lambda_3$, i.e., the weight of the recovery-time, efficiently. Note, that for $\lambda_3\ll1$ and large $T_o$, the proposed algorithm reduces to behaving like the conventional convergence-based algorithms, thus representing a generalization to them. The detailed steps of the resilience-aware alternating optimization are illustrated in Algorithm \ref{alg}.

\begin{algorithm}
\footnotesize
\caption{Resilience-aware Alternating Optimization}\label{alg}
\begin{tikzpicture}[>={Latex[length=5pt]}]\label{alg:AltOpt}
	\footnotesize
	\tikzset{myRect/.style={draw,rectangle,minimum width=0.5cm, minimum height=0.25cm,align=center}}
	\tikzset{myDiam/.style={draw,diamond,aspect=1.5,minimum width=1.25cm, minimum height=0.9cm,align=center}}
	\tikzset{myArrow/.style={->,draw,line width=0.25m}}
	\tikzset{myDot/.style={draw,minimum size=0.1,circle,fill,scale=0.15}}
	
	\node[align=center](input) at (-1,-1.85){ Input:\\ $\tilde{\vect{w}},\tilde{\vect{v}},\tilde{\vect{\kappa}},$\\ $T_w,T_v,o,\tau$ };
	
	\node[myRect](initLam) at ($(input)+(0,2.45)$){Create\\$\tilde{\mat{\Lambda}}_z^o$};

	\node[myDiam] (o1) at (0,0){};
	\node[] (o1txt) at (o1){$o=w$};
	\node[myRect] (P2) at ($(o1)+(2.625,0.3)$){$\hat{\vect{\Lambda}}_{z+1}^w \leftarrow$ solve (P2.1)};
	\node[myRect] (P3) at ($(o1)+(2.625,-0.3)$){$\hat{\vect{\Lambda}}_{z+1}^v \leftarrow$ solve (P3.1)};
	\node[myDiam] (oT) at (6,0){};
	\node[] (oTText) at (oT){$T{<}\:T_o$};
	\node[myDiam] (o2) at (4.555,-1.5){};
	\node[] (o2txt) at (o2){$o=w$};
	\node[myRect,align=center] (Lambw) at ($(o2)+(-2.45,0.45)$){$[\tilde{\vect{w}}^T, \tilde{\vect{\kappa}}^T]^T \leftarrow \hat{\mat{\Lambda}}_z^w$ ,\\$\tilde{\mat{\Lambda}}_z^v \leftarrow [\tilde{\vect{v}}^T, \tilde{\vect{\kappa}}^T]^T $};
	\node[myRect,align=center] (Lambv) at ($(o2)+(-2.45,-0.45)$){$[\tilde{\vect{v}}^T, \tilde{\vect{\kappa}}^T]^T \leftarrow \hat{\mat{\Lambda}}_z^v$ \\ $\tilde{\mat{\Lambda}}_z^w \leftarrow [\tilde{\vect{w}}^T, \tilde{\vect{\kappa}}^T]^T$};
	
	\node[myDiam](Psi) at ($(oT)+(-0.45,-1.025)$){};
	\node(Psitxt) at (Psi) {$\Uppsi<\tau$};

	\draw[->] (input.north) |- ($(input.north)!0.25!(initLam.240)$) -|node[pos=0.535,anchor=210]{$\phantom{.}_{T{\leftarrow 0}}$}node[pos=0.655,anchor=210]{$\phantom{.}_{z{\leftarrow 0}}$} (initLam.240);
	\draw[->] (initLam.south) |- (o1.west);

	\draw[->] (o1.east) --node[myDot,pos=1]{} ($(o1.east)+(0.1,0)$) |-node[pos=0.65,above]{\tiny$\mathsf{T}$} (P2.west) ;
	\draw[->] (o1.east) -- ($(o1.east)+(0.1,0)$)  |-node[pos=0.65,above]{\tiny$\mathsf{F}$} (P3.west) ;
	
	\node[myDot] (incrementDot) at ($(oT.west)-(1.15,0)$){};
    \draw[->] (P2.east) -| (incrementDot);
	\draw[->] (P3.east) -| (incrementDot);
	
	\draw[->] (incrementDot) --node[pos=0.475,above]{$\phantom{.}_{z\,{\leftarrow}{z+1}}$}node[pos=0.475,below]{$\phantom{.}_{T{\leftarrow}{T+1}}$} (oT.west);
	
	\draw[->] (oT.north) |- node[pos=0.1,right]{\tiny$\mathsf{T}$} node[pos=0.65,below]{$\tilde{\mat{\Lambda}}_z^o \leftarrow \hat{\mat{\Lambda}}_z^o$}  ($($(oT.north)!0.5!(o1.north)$) + (0,0.4) $) -| (o1.north);
	
	\draw[->] (oT.south) -| node[pos=0.15,anchor=-35]{\tiny$\mathsf{F}$}  (o2.north); 
	
	\draw[->] (o2.west) --node[myDot,pos=1]{} ($(o2.west)-(0.125,0)$) |-node[pos=0.65,above]{\tiny$\mathsf{T}$} (Lambw.east) ;
	\draw[->] (o2.west) -- ($(o2.west)-(0.125,0)$)  |-node[pos=0.65,above]{\tiny$\mathsf{F}$} (Lambv.east) ;
	
	\node[myDot] (reassignDot) at  ($(o2)-(4.39,0)$) {};
	\draw[->] (Lambw.west) -|node[pos=0.25,above]{$\phantom{.}_{o\leftarrow v}$} (reassignDot) ;
    \draw[->] (Lambv.west) -|node[pos=0.25,above]{$\phantom{.}_{o\leftarrow w}$} (reassignDot) ;

    \draw[->] (reassignDot) -|node[pos=0.65,anchor=-25]{$\phantom{.}_{T{\leftarrow} 0}$} (o1.south);
	
	\draw[->] (oT.east) |-node[pos=0.85,anchor=125,align=center](a){} ($(Psi.east)+(0.1,0)$) -- (Psi.east); 
	
	\node[align=center] at ($(a)+(0,-0.35)$){Output:\\$\hat{\vect{w}}_z,\hat{\vect{v}}_z$};
	
	\draw[->] (Psi.south) --node[pos=0.25,right]{\tiny$\mathsf{T}$} ($(Psi.south)+(0,-0.4)$);
	
	\node at ($(Psi.south)+(0,-0.65)$) {Stop};
	
\end{tikzpicture}
\end{algorithm}

%% file: sections/numerical_results.tex
\section{Numerical results}
In this section, we numerically evaluate the performance of an \ac{RIS} as a resilience method. To this end we assume a cell-free \ac{MIMO} system with $N=3$ \acp{AP}, each of which equipped with $L=14$ antennas. We assume the $K=14$ single-antenna users, as well as the \acp{AP}, to be distributed randomly within an area of operation, which spans $[-250,250] \times [-250,250] \text{m}^2$. The \ac{RIS} is positioned in the center of this area and is assumed to be composed of $M = 196$ reflecting elements, deployed in a quadratic grid with $\frac{\lambda}{4}$ spacing, where $\lambda = 0.1$m is the wavelength. Hence, we employ the correlated channel model introduced in \cite{corrBj}, where the average attenuation intensity is modeled after \cite[Eq.(23)]{BjonAtten}. For the direct links we assume Rayleigh fading channels with log-normal shadowing with 8dB standard deviation. Further, we assume a bandwidth of $B=10 \text{MHz}$, a noise power of $-100$dBm, a maximum transmit power of $P_n^\mathsf{Max} = 40$dBm per \ac{AP} and each user to require a \ac{QoS} of $r_k^\mathsf{des}=12$Mbps.
We define the occurrence of an outage as an event, where each individual direct link between \ac{AP} $n$ and user $k$ has a 12\% probability to be subject to a complete blockage. Note that the \ac{RIS} has been positioned in a way that the RIS-assisted links are exempt from blockages. In addition, we assume that rate adaption as a resilience mechanism is utilized right after every outage occurs \cite[M1]{RobertRes}. To this end, Problem (\ref{Prob2.1}) is immediately solved with fixed beamformers after occurrence of any outage. The network operator's desired recovery time is set to $T_0 = 0$ms.

\subsection{Convergence Behaviour}
\begin{figure}
	\centering
\includegraphics[width=1.0\linewidth]{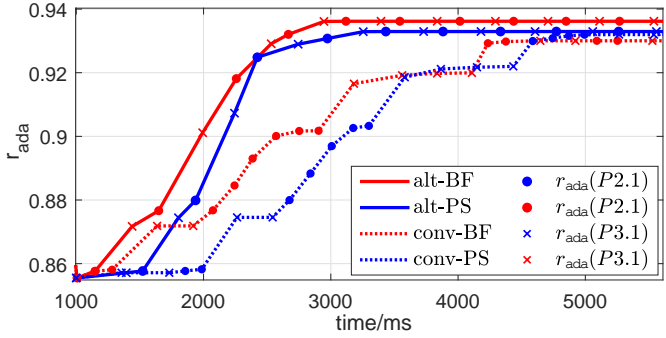}
\caption{Convergence behaviour of the alternating (alt) and convergence (conv) based approaches when initializing with either the beamforming (BF) or phase-shifting (PS) sub-problem after an outage occurs at $t_0=1000$ms.}
\label{altVSconv}
\end{figure}

First we study the effect of the amount of fixed iterations $T_o$ employed in Algorithm \ref{alg}. To this end, we define two approaches: 1) an alternating approach (alt), in which we set $T_o=1, \forall o\in\{w,v\}$ and 2) a convergence-based approach (conv), in which each-subproblem is optimized until convergence. At this point, it also becomes important to decide which sub-problem is solved first after an outage occurs. Thus, we also compare the performance of both approaches above when initializing with either the beamforming problem $o=w$ (alt-BF, conv-BF) or the phase shifting problem $o=v$ (alt-PS, conv-PS) after an outage has occurred. Fig. \ref{altVSconv} depicts an outage, occuring at $t_0=1000$ms, and the changes in $r_{ada}$ of the different approaches over the time required to solve the respective sub-problems. Note that the slope between any two points of a curve represents the quality-time trade-off between those points.

It becomes apparent that both red curves, representing the approaches starting with the beamforming sub-problem, perform better right after the outage has occurred. The rationale behind this behaviour lies in the fact that the beamformer of any \ac{AP}-link that is affected by an outage should be redirected to the \ac{RIS} first, before reflecting any incoming signals at the \ac{RIS} to the users. Fig. \ref{altVSconv} also shows that the alternating approaches are performing better with regards to the quality-time ratio between each iteration than the convergence-based ones. In addition, the alternating approach starting with the beamforming sub-problem (alt-BF) not only performs the best considering the quality-time ratios but also finds the best solution after convergence of Algorithm \ref{alg}, i.e., when the time-to-recovery does not play a role.
\subsection{Impact and Operating Modes of RIS as Resilience Method}
\begin{figure}
	\centering
\includegraphics[width=1\linewidth]{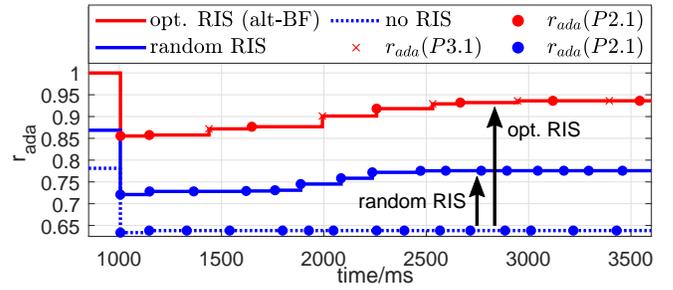}
\caption{Adaption of the same system when 1) optimizing the \ac{RIS} (alt-BF), 2) keeping the \ac{RIS} random and 3) without \ac{RIS} after an outage occurs at $t_0=1000$ms.}
\label{RISbenef}\vspace{-0cm}
\end{figure}

To illustrate the impact of the \ac{RIS} on the resilience scope of the beamforming sub-problem, Fig. \ref{RISbenef} compares a conventional system (without \ac{RIS}) with the \ac{RIS}-assisted system. For the \ac{RIS}-assisted case, we distinguish between two modes: random phase shifting and optimized phase shifting. The figure shows that even before the outage occurs at $t_0=1000$ms, the conventional system without \ac{RIS} performs worse that the \ac{RIS}-assisted ones. Adding the \ac{RIS} with random phase shifters already improves the performance of the system by around 10\%, while optimized phase shifts increase the performance by around 28\% and enable the system to satisfy the \ac{QoS} for every user ($r=1$).
After the outage occurs, the system without the \ac{RIS} adjusts the beamformers by solving problem (\ref{Prob2.1}) and converges after only one iteration. This is due to the fact that the available resources were already too scarce to satisfy every user's \ac{QoS}. Thus, the system can only react by reallocating the power of those beamformers, which are subjected to a blockage, resulting in a performance loss of 19\%.

By adding a randomly configured \ac{RIS} to the system, the figure shows that a better resilience behaviour can be obtained, due to the additional paths provided by the \ac{RIS}. This enables the system to not only reallocate the power, but also re-adjust the beamformers towards the \ac{RIS}. This degrades the performance by only 10\% as a result of the extended resilience scope provided by the additional \ac{RIS} reflections. Including the phase shifters at the \ac{RIS} into the optimization process (alt-BF), the system is able to re-adjust the beamformers and the phase shifters, which results in a degradation of merely 6\%.
\begin{figure}
	\centering
	\includegraphics[width=1\linewidth]{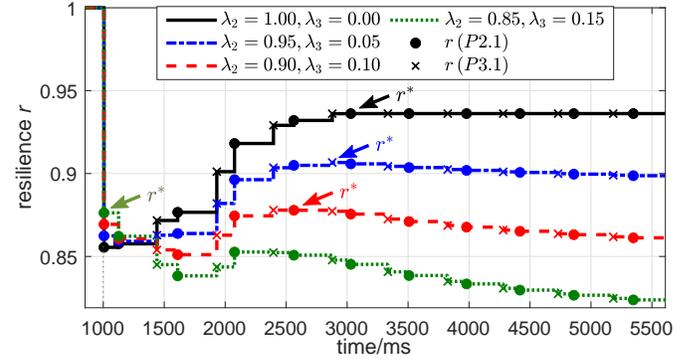}
	\caption{Resilience behavior for different $\lambda_i$ values for the \ac{RIS}-assisted system using the (alt-BF) algorithm after an outage occurs at $t_0=1000$ms. The arrows show the iteration, which has the optimal quality-time trade-off for the respective configurations.}
	\label{lambVals}\vspace{-0cm}
\end{figure}
\subsection{Resilience for Different Requirements $\lambda_i$}
In this subsection, we bridge the gap towards the resilience metric proposed in (\ref{eq:r}). To this end, Fig. \ref{lambVals} shows the resilience $r$ for different weights $\lambda_2$ and $\lambda_3$, i.e., preferred quality-time trade-offs of the system. The black curve in the figure depicts the setting $\lambda_2=1$ and $\lambda_3=0$. With this  setting, the resilience does not depend on the required time-to-recovery $r_{rec}$ and consequently coincides with the adaption of the alt-BF scheme in Fig. \ref{RISbenef}. Hence, for this case, the best performing iteration is obtained after convergence. However, by slightly increasing the impact of $r_{rec}$, Fig. \ref{lambVals} shows that after the outage, the remaining curves are characterized by a decline followed by a peak. With increasing values of $\lambda_3$, the peak is not only closer to the outage, but also less pronounced. This is due to the weakened importance of the quality of the solution, represented by the inherent decline of the weight $\lambda_2$. Interestingly, the green curve shows that the time-to-recovery can also consist of values, at which the time required for one iteration can become a limiting factor. For these cases, the rate adaption becomes the most efficient resilience method due to its fast adaption time.
\begin{figure}
	\centering
	\includegraphics[width=1\linewidth]{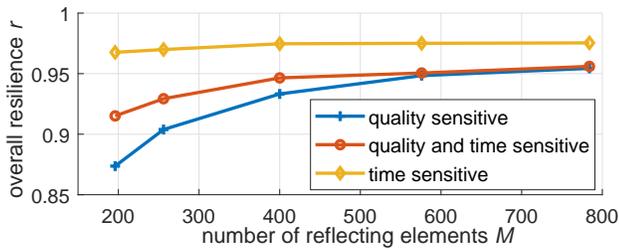}
	\caption{Overall resilience of the best performing iterations for the quality-sensitive, time-sensitive and quality-and-time-sensitive setups over the number of reflecting elements $M$.}
	\label{RoverM}\vspace{-0.275cm}
\end{figure}
\begin{figure}
	\centering
	\includegraphics[width=\linewidth]{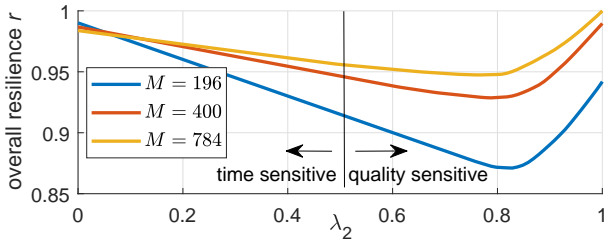}
	\caption{Overall resilience of the best performing iterations for different numbers of reflecting elements $M$ over all possible weight combinations of $\lambda_i$.}
	\label{MvsLamb}\vspace{-0.275cm}
\end{figure}
\subsection{Impact of the Number of Reflecting Elements}
By adding more reflecting elements to the system, the amount of RIS links increases, which improves the quality of the adaption $r_\textsf{ada}$. However, this improved quality comes at the cost of increased computational complexity, which is detrimental for the time that is needed to determine the optimal configuration. In the context of resilience, Fig. \ref{RoverM} shows how the number of reflecting elements $M$ directly influences the resilience for different weight configurations of $\lambda_i$. Note that the \textit{overall} resilience value $r$ is obtained by determining the best performing iteration as depicted by the arrows in Fig. \ref{lambVals}. Interestingly, all setups - one that favours the time-to-recovery (time sensitive, $\lambda_2 = 0.85$, $\lambda_3 = 0.15$), one that favours the quality of the solution (quality sensitive, $\lambda_2 = 0.15$, $\lambda_3 = 0.85$) and one that favours both equally (quality and time sensitive, $\lambda_2 = \lambda_3 = 0.5$) -  benefit from having more reflecting elements in the system even when the time-to-recovery is detrimental. This is due to the fact, that in the time-sensitive setups only the first iteration, i.e., the beamforming solution of (\ref{Prob2.1}) is fast enough to contribute to the resilience performance. Even though the \ac{RIS} is not optimized in this case, more reflecting elements still improve the resilience as they support the resilience scope of the beamforming problem by passively providing additional links.

In order to gain a more detailed insight of how the number of \ac{RIS} elements influences the resilience performance, Fig. \ref{MvsLamb} plots the overall resilience $r$ for different \ac{RIS} sizes over all possible weight combinations of $\lambda_i$. The figure shows that adding more \ac{RIS} elements to the system is beneficial for all but the most time-sensitive setups, i.e., $\lambda_2<0.05$, as for these cases more \ac{RIS} elements result in a higher complexity, therefore more time, when calculating the rate adaption. Surprisingly, Fig. \ref{MvsLamb} also shows a distinct minimum for all \ac{RIS} sizes for $\lambda_2\approx0.81$, at which the resilience-aware alternating optimization approach seems to perform the worst. However, it is also at this point, where adding additional reflecting elements to the system results in the highest gains.

%% file: sections/conclusion.tex
\section{Conclusion}\label{ch:conc}
This work studies the benefits that can be obtained by utilizing the \ac{RIS} to act as a resilience mechanism in order to counteract outages effectively. Due to its passive nature, even unoptimized \ac{RIS}-assisted reflections are able to contribute to the improvement of the system's resilience. By including the \ac{RIS} into the optimization process with the proposed resilience-aware alternating optimization, the benefits of the \ac{RIS} become even more pronounced. Somewhat surprisingly, it is shown that adding more reflecting elements to the system results in a better resilience performance regardless of the system requirements. Except for the most time-sensitive systems, the gain in quality of the adaption due to the additional links outweigh the fact that adding more elements to the system increases its complexity, and thus its time-to-recovery. We conclude that utilizing the \ac{RIS} for resilience-relevant applications is shown to be a promising avenue for enabling reliable and fault-tolerant future wireless networks. 

%% file: content/laststuff.tex
%




%







%% file: content/acronyms.tex
\begin{acronym}
\setlength{\itemsep}{0.1em}
\acro{AP}{access point}
\acro{AF}{amplify-and-forward}
\acro{AWGN}{additive white Gaussian noise}
\acro{B5G}{Beyond 5G}
\acro{BS}{base station}
\acro{CB}{coherence block}
\acro{CE}{channel estimation}
\acro{C-RAN}{Cloud Radio Access Network}
\acro{CMD}{common message decoding}
\acro{CP}{central processor}
\acro{CSI}{channel state information}
\acro{CRLB}{Cramér-Rao lower bound}
\acro{D2D}{device-to-device}
\acro{DC}{difference-of-convex}
\acro{DFT}{discrete Fourier transformation}
\acro{DL}{downlink}
\acro{GDoF}{generalized degrees of freedom}
\acro{IC}{interference channel}
\acro{i.i.d.}{independent and identically distributed}
\acro{IRS}{intelligent reflecting surface}
\acro{IoT}{Internet of Things}
\acro{LoS}{line-of-sight}
\acro{LSF}{large scale fading}
\acro{M2M}{Machine to Machine}
\acro{MISO}{multiple-input and single-output}
\acro{MIMO}{multiple-input and multiple-output}
\acro{MRT}{maximum ratio transmission}
\acro{MRC}{maximum ratio combining}
\acro{MSE}{mean square error}
\acro{NOMA}{non-orthogonal multiple access}
\acro{NLoS}{non-line-of-sight}
\acro{PSD}{positive semidefinite}
\acro{QCQP}{quadratically constrained quadratic programming}
\acro{QoS}{quality-of-service}
\acro{RF}{radio frequency}
\acro{RC}{reflect coefficient}
\acro{RIS}{reconfigurable intelligent surface}
\acro{RS-CMD}{rate splitting and common message decoding}
\acro{RSMA}{rate-splitting multiple access}
\acro{RS}{rate splitting}
\acro{SCA}{successive convex approximation}
\acro{SDP}{semidefinite programming}
\acro{SDR}{semidefinite relaxation}
\acro{SIC}{successive interference cancellation}
\acro{SINR}{signal-to-interference-plus-noise ratio}
\acro{SOCP}{second-order cone program}
\acro{SVD}{singular value decomposition }
\acro{TIN}{treating interference as noise}
\acro{TDD}{time-division duplexing}
\acro{TSM}{topological signal management}
\acro{UHDV}{Ultra High Definition Video}
\acro{UL}{uplink}

\acro{AF}{amplify-and-forward}
\acro{AWGN}{additive white Gaussian noise}
\acro{B5G}{Beyond 5G}
\acro{BS}{base station}
\acro{C-RAN}{Cloud Radio Access Network}
\acro{CSI}{channel state information}
\acro{CMD}{common-message-decoding}
\acro{CM}{common-message}
\acro{CoMP}{coordinated multi-point}
\acro{CP}{central processor}
\acro{D2D}{device-to-device}
\acro{DC}{difference-of-convex}
\acro{EE}{energy efficiency}
\acro{IC}{interference channel}
\acro{i.i.d.}{independent and identically distributed}
\acro{IRS}{intelligent reflecting surface}
\acro{IoT}{Internet of Things}
\acro{LoS}{line-of-sight}
\acro{LoSC}{level of supportive connectivity}
\acro{M2M}{Machine to Machine}
\acro{NOMA}{non-orthogonal multiple access}
\acro{MISO}{multiple-input and single-output}
\acro{MIMO}{multiple-input and multiple-output}
\acro{MMSE}{minimum mean squared error}
\acro{MRT}{maximum ratio transmission}
\acro{MRC}{maximum ratio combining}
\acro{NLoS}{non-line-of-sight}
\acro{PA}{power amplifier}
\acro{PSD}{positive semidefinite}
\acro{QCQP}{quadratically constrained quadratic programming}
\acro{QoS}{quality-of-service}
\acro{RF}{radio frequency}
\acro{RRU}{remote radio unit}
\acro{RS-CMD}{rate splitting and common message decoding}
\acro{RS}{rate splitting}
\acro{SDP}{semidefinite programming}
\acro{SDR}{semidefinite relaxation}
\acro{SIC}{successive interference cancellation}
\acro{SCA}{successive convex approximation}
\acro{SINR}{signal-to-interference-plus-noise ratio}
\acro{SOCP}{second-order cone program}
\acro{SVD}{singular value decomposition }
\acro{TP}{transition point}
\acro{TIN}{treating interference as noise}
\acro{UHDV}{Ultra High Definition Video}
\acro{URLLC}{ultra reliable and low-latency communication}
\acro{LoSC}{level of supportive connectivity}
\end{acronym}